\documentclass[%
reprint,
superscriptaddress,
fnobibnotes,
 amsmath,amssymb,
 aps,
longbibliography,
floatfix,
]{revtex4-1}

\usepackage{graphicx}
\usepackage{dcolumn}
\usepackage{bm}
\usepackage{svg}
\usepackage{xcolor}
\usepackage[colorlinks=true,linkcolor=purple,citecolor=purple]{hyperref}%
\usepackage{enumitem}
\usepackage{siunitx}
\usepackage[normalem]{ulem}
\usepackage{physics}
\usepackage[pagewise]{lineno}
\usepackage{multirow}

\newcommand{\zlatko}[1]{ { \color{green} (ZM: {#1}) }}

\makeatletter
\renewcommand{\fnum@figure}{\textbf{Fig.~\thefigure}}
\makeatother

\begin{document}

\title{Machine Learning for Practical Quantum Error Mitigation}

\author{Haoran Liao}
\thanks{These authors contributed equally}
\affiliation{%
IBM Quantum, IBM T.J. Watson Research Center, Yorktown Heights, NY 10598, USA
}%
\affiliation{Department of Physics, University of California, Berkeley, CA 94720, USA}

\author{Derek S. Wang}
\thanks{These authors contributed equally}
\affiliation{%
IBM Quantum, IBM T.J. Watson Research Center, Yorktown Heights, NY 10598, USA
}%

\author{Iskandar Sitdikov}
\affiliation{%
IBM Quantum, IBM T.J. Watson Research Center, Yorktown Heights, NY 10598, USA
}%

\author{Ciro Salcedo}
\affiliation{%
IBM Quantum, IBM T.J. Watson Research Center, Yorktown Heights, NY 10598, USA
}

\author{Alireza Seif}
\affiliation{%
IBM Quantum, IBM T.J. Watson Research Center, Yorktown Heights, NY 10598, USA
}

\author{Zlatko K. Minev}
\email{haoran.liao@berkeley.edu\\zlatko.minev@ibm.com}
\affiliation{%
IBM Quantum, IBM T.J. Watson Research Center, Yorktown Heights, NY 10598, USA
}

\begin{abstract}
Quantum computers progress toward outperforming classical supercomputers, but quantum errors remain their primary obstacle. The key to overcoming errors on near-term devices has emerged through the field of quantum error mitigation, enabling improved accuracy at the cost of additional run time. 
Here, through experiments on state-of-the-art quantum computers using up to 100 qubits, we demonstrate that without sacrificing accuracy machine learning for quantum error mitigation (ML-QEM) drastically reduces the cost of mitigation. We benchmark ML-QEM using a variety of machine learning models---linear regression, random forests, multi-layer perceptrons, and graph neural networks---on diverse classes of quantum circuits, over increasingly complex device-noise profiles,  under interpolation and extrapolation, and in both numerics and experiments. These tests employ the popular digital zero-noise extrapolation method as an added reference. Finally, we propose a path toward scalable mitigation by using ML-QEM to mimic traditional mitigation methods with superior runtime efficiency. Our results show that classical machine learning can extend the reach and practicality of quantum error mitigation by reducing its overheads and highlight its broader potential for practical quantum computations. 
\end{abstract}

\maketitle

\section{\label{sec:intro}Introduction}
Quantum computers hold the promise of substantial advantages over their classical counterparts, with speedups ranging from polynomial to exponential~\cite{Biamonte2017a, Daley2022}. 
However, realizing these advantages in practice is hindered by unavoidable errors in the physical quantum devices.
Achieving reduced error rates and increasing qubit numbers will in principle  ultimately allow fault-tolerant quantum error correction to overcome these errors \cite{Campbell2017}.
While this goal remains out of reach, quantum error mitigation (QEM) strategies have been developed to harness imperfect quantum computers to produce near noise-free results despite the presence of unmonitored errors \cite{Bravyi2022,GoogleQEMReview2022, Kandala2019, Berg2022, Kim2023, Youngseok2023, Daley2022, Youngseok2023}.  
These strategies are not just a temporary fix, but are an essential step towards achieving near-term quantum utility and establishing a path to outperform classical supercomputers~\cite{Daley2022, Youngseok2023}.

The main challenge to employing QEM in practice lies in devising schemes that yield accurate results without excessive runtime overheads. For context, quantum error correction relies on overheads in qubit counts and real-time monitoring  to eliminate errors in each run of a circuit. In contrast, QEM obviates the need for these overheads but at the cost of increased algorithmic run time. Instead, QEM produces an estimator for the noise-free expectation values of a target circuit by employing an ensemble of many noisy quantum circuits. 
For example, in the cornerstone QEM approach known as zero-noise extrapolation (ZNE) \cite{Temme2016, Li2017, tsubouchi2023universal}, an input circuit is recompiled into multiple circuits that are logically equivalent but each with an increased expected number of errors. By analyzing the dependence of the measured expectation values for each noisy circuit, one can estimate the `zero-noise', ideal expectation value of the original circuit. While ZNE does not yield an unbiased estimator, other QEM methods, such as probabilistic error cancellation (PEC)  \cite{Temme2016, Li2017, Berg2022} come fortified with rigorous theoretical guarantees and sampling complexity bounds. Unfortunately, it is believed that QEM methods demand exponential sampling overheads for arbitrarily high accuracies\cite{Quek2022, Takagi2022, tsubouchi2023universal}---making them challenging to implement at increasing scales of interest. 
Thus, the quest for QEM methods that balance scalability, cost-effectiveness, and generality remains at the forefront of quantum computing research.

Emerging at the crossroads of quantum mechanics and statistical learning, machine learning for quantum error mitigation (ML-QEM) presents a promising avenue where statistical models are trained to derive mitigated expectation values from noisy counterparts executed on quantum computers. 
Could such ML-QEM methods offer valuable improvement in accuracy or runtime efficiency in practice? 

\begin{figure*}[!thbp]
\centering
\includegraphics[width=0.9\textwidth]{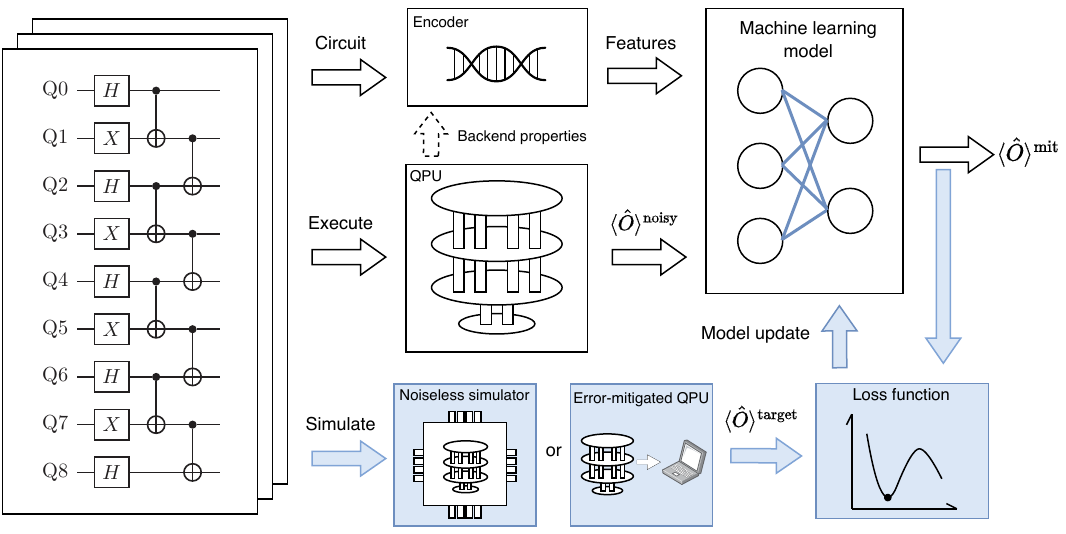}
\caption{
\textbf{Execution and training for tractable and intractable circuits with ML-QEM.} 
A quantum circuit (left) is passed to an encoder (top) that creates a feature set for the ML model (right) based on the circuit and the quantum processor unit (QPU) targeted for execution. The model and features are readily replaceable. The executed noisy expectation values $\langle \hat O \rangle^\mathrm{noisy}$ (middle) serve as the input to the model whose aim is to predict their noise-free value  $\langle \hat O \rangle^\mathrm{mit}$. 
To achieve this, the model is trained beforehand (bottom, blue highlighted path) against target values $\langle \hat O \rangle^\mathrm{target}$ of example circuits. These are obtained either using noiseless simulations in the case of small-scale, tractable circuits or using the noisy QPU in conjunction with a conventional error mitigation strategy in the case of large-scale, intractable circuits. The training minimizes the loss function, typically the mean square error. The trained model operates without the need for additional mitigation circuits, thus reducing runtime overheads.
}
\label{fig:workflow}
\end{figure*}

Theoretically, a successful ML-QEM strategy could learn the effects of noise on the execution of a quantum circuit during training and subsequently mitigate the noisy classical data from the quantum processor, such as quantum expectation values, when deployed.
This eliminates the need for extra mitigation circuits during algorithm execution, reducing the algorithmic runtime overhead compared to conventional QEM.
In practice, however, quantum noise can be intricately complex, high-dimensional, and can even drift stochastically over time.

Early explorations of ML-QEM ideas have shown signs of promise, even for some complex noise profiles~\cite{Kim2020, Czarnik2021, Czarnik2022, Bennewitz2022, Tirthak2021, PRXQuantum.2.040330}, but it remains unclear if ML-QEM can perform in practice in quantum computations on hardware or at scale. For instance, it is unclear whether a given ML-QEM method can be used across different device noise profiles, diverse circuit classes, and large quantum circuit volumes beyond the limits of classical simulation.
To date, there has not been a systematic study comparing different traditional methods and statistical models for QEM on equal footing under practical scenarios across a variety of relevant quantum computational tasks.

In this article, we present a general framework for ML-QEM, offering superior runtime efficiency over conventional error mitigation techniques. Our study encompasses a broad spectrum of simple to complex machine learning models, including the previously proposed linear regression and multi-layer perceptron. We further propose two new models, random forests regression and graph neural networks. Among these, we identify random forests as the consistently best-performing model. We evaluate the performance of all four models in diverse realistic scenarios. We consider a range of circuit classes (random circuits and Trotterized Ising dynamics) and increasingly complex noise models in simulations (including incoherent, coherent, and readout errors).
Additionally, we explore the advantages of ML-QEM methods over traditional popular approaches in common-use application scenarios, such as to mitigate unseen Pauli observables in a tomography-like setting or to enhance variational quantum-classical tasks. 

Our analysis reveals that ML-QEM methods, particularly random forest, not only rival but can even surpass the overall performance of the benchmarked popular digital zero-noise extrapolation (ZNE) method.
Even in the most conservative setting, ML-QEM demonstrates more than a 2-fold reduction in runtime overhead, 
addressing primary bottleneck of error mitigation and a fundamentally challenging problem with a substantial leap in efficiency. 
Most conservativeley, this translates into at least halving experiment durations---e.g., cutting an 80-hour experiment to just 40 hours~\cite{Shtanko2023}---drastically reducing operational costs and doubling the regime of accesible experiments. 
With successful experiments on IBM quantum computers for quantum circuits with up to 100 qubits and two-qubit gate depth of 40, we chart a viable path toward scalable error mitigation. By mimicking traditional methods while achieving superior runtime efficiency, ML-QEM underscores the transformative potential of leveraging classical machine learning to enhance quantum data processing~\cite{Huang_2021, Huang2022}.

\section{\label{sec:results}Results}

The ML-QEM workflow, illustrated in Fig.~\ref{fig:workflow}, is tailored for specific classes of quantum circuits. Within this framework, we train an ML model to estimate noise-free expectation values from noisy values retrieved from a quantum processing unit (QPU), incorporating both QPU and circuit features. This approach is crucial because the quantum circuit's output is typically intractable, making it challenging for the ML model to learn the results independent of the QPU noisy values. A comprehensive overview of the training set, encoded features, and ML models is provided in the Methods section. Notably, during runtime, the ML-QEM model derives mitigated expectation values directly from the noisy data, eliminating the need for supplementary mitigation circuits and thus substantially cutting overheads.

We begin by assessing performance on small-scale circuits via numerical simulations under realistic noise models before transitioning to real-world noisy hardware validation. Subsequently, we introduce a scalable ML-QEM approach for large circuits, exemplified through experiments on 100-qubit circuits containing up to 1,980 CNOT gates. In this regime, the calculations are beyond simple brute force numerical techniques, and serve as a test-bed for intractable circuits.

\subsection{\label{ssec:comparison}Performance Comparison at Tractable Scales}

First, we compare a diverse array of ML-QEM models to establish which ML methods may be best suited for quantum error mitigaiton (QEM). We focus on four statistical models: linear regression with ordinary least squares (OLS), random forests regression (RF), multi-layer perceptrons (MLP), and graph neural networks (GNN). Given the inherent generally non-linear relationship between noisy and ideal expectation values (as discussed in App.~\ref{app:depolarizing_noise}), the significance of non-linear models is emphasized. 

We evaluate the performance of these methods for two classes of circuits: random circuits and Trotterized dynamics of the 1D Ising spin chain on small-scale simulations. These two classes of circuits 
bear distinct two-qubit gate arrangements, allowing us to gain knowledge about the performance of the ML-QEM on the two extremes of the spectrum in terms of circuit structures. 

In the study of Trotterized circuits, we also study in simulations the performance of the methods in the absence and presence of readout error or coherent noise, in addition to incoherent noise.

\subsubsection{Random Circuits}\label{ssec:random_cricuits}

\begin{figure}[!ht]
\centering
\includegraphics[width=0.47\textwidth]{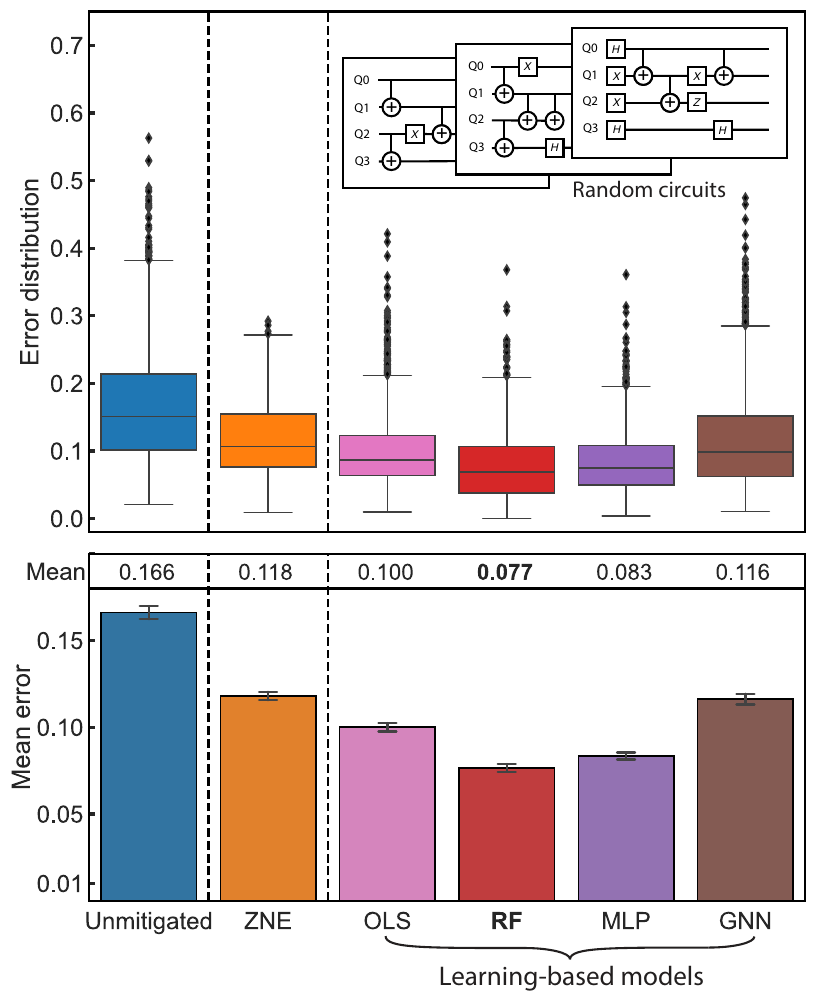}
\caption{
\textbf{Accuracy of QEM and ML-QEM in random circuits.}
Top: Error distribution for unmitigated and mitigated Pauli-$Z$ expectation values. 
 Mitigation is performed using either a reference QEM method, digital zero-noise extrapolation (ZNE), or one of four ML-QEM models (explained in text). 
Inset: Example random circuits.
Noisy execution is numerically simulated using a noise model derived from IBM QPU \texttt{Lima}, \texttt{FakeLima}. The error is defined as the $L_2$ distance between the vector of all ideal and noisy single-qubit expectations $\langle \hat{Z}_i \rangle$; i.e., $\Vert \langle \hat{Z} \rangle - \langle  \hat{Z} \rangle_{\mathrm{ideal}} \Vert_2$. 
Distribution is over $n=2{,}000$ four-qubit random circuits, with two-qubit-gate depths sampled up to 18 with a step size of 2. 
The horizontal lower boundary, center line, and upper boundary of a colored box indicate the first, second, and third quantile, respectively. The horizontal lines outside indicate the $1.5\times$(interquartile range) from the nearest hinge. Black dots are outliers.
Bottom: Mean $L_2$ error over the $n=2{,}000$ circuits for each method (using data from the top). The value is provided above each column. The errorbar represents the $95\%$ confidence intervals derived from $1{,}000$ bootstrap re-sampling. 
}
\label{fig:random_circuits}
\end{figure}

In the first experiment, we benchmark the performance of the protocol on small-scale unstructured circuits. To ensure that the circuits encompass a broad spectrum of complexities, we generate a diverse set of four-qubit random circuits with varying two-qubit gate depths, up to a maximum of $18$ with a step size of 2, as shown in the inset of Fig.~\ref{fig:random_circuits}. Per two-qubit depth, there are $500$ random training circuits and $200$ random test circuits that are generated by the same sampling procedure. For each circuit, we carry out simulations on IBM's \texttt{FakeLima} backend, which emulates the incoherent noise present in the real quantum computer, the \texttt{ibmq\_lima} device. While these quantum devices generally have coherent errors as well, they can be suppressed through a combination of e.g., dynamical decoupling~\cite{Ezzell2022, Pokharel2022, Seif_2024} and randomized compiling~\cite{Bennett1996,  Wallman2015,Hashim2021,Berg2022}. Specific types of noise include incoherent gate errors, qubit decoherence, and readout errors. We train the ML-QEM models to mitigate the noisy expectation values of the four single-qubit $\hat{Z}_i$ observables. As a benchmark, we also compare mitigated expectation values from digital ZNE (see Sec.~\ref{sec:zne_configuration} for details). In Fig.~\ref{fig:random_circuits}, we show the error (between the mitigated expectation values and the ideal ones) distribution of ZNE and ML-QEM with each of the four machine learning models on the top panel and the bootstrap mean errors in the bottom panel. We observe that RF consistently outperforms the other ML-QEM models. Notably, all ML-QEM models exhibit competitive performance in comparison to digital ZNE, despite that the runtime overhead for digital ZNE is twice as much.

\subsubsection{Trotterized 1D Transverse-field Ising Model}\label{sec:comparison_ising}

\begin{figure*}
\centering
\includegraphics[width=0.97
\textwidth]{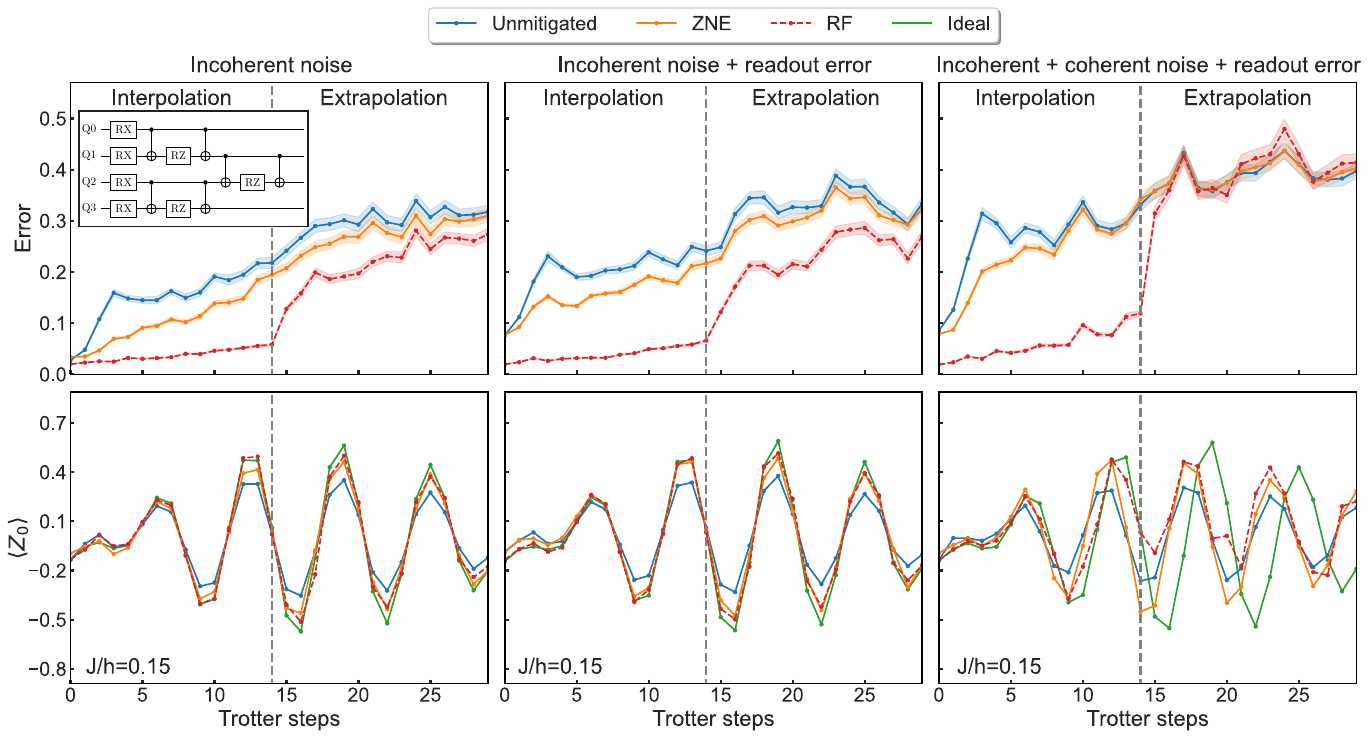}
\caption{
\textbf{Mitigation accuracy under i)  complexity of quantum noise and ii) ML-QEM interpolation and extrapolation for Trotter circuits.}
Top row: Average error performance on Trotter circuits (top-left inset) representing the time evolution dynamics of a four-site, 1D, transverse-field Ising model in numerical simulations. A Trotter step comprises four layers of CNOT gates (inset). Vertical dashed line separates experiments in the ML-QEM interpolation regime (left) from the extrapolation regime (right). The 3 curves represent the performance of the highest-performing ML-QEM method, the ZNE method, and the unmitigated experiments. They are averaged over 300 circuits, each with a randomly chosen Pauli measurement bases. The data is for all four weight-one expectations $\langle \hat{P}_i \rangle$. The error is defined as $L_2$ distance from the ideal expectations, $\Vert \langle \hat{P} \rangle - \langle  \hat{P} \rangle_{\mathrm{ideal}} \Vert_2$, as also defined for the remainder of figures. From the left to right, the complexity of the device noise model increases to include additional realistic noise types. Coherent errors are introduced on CNOT gates. 
Bottom row: 
Corresponding typical data of the error-mitigated expectation values of the  $\langle \hat{Z}_0 \rangle$ Trotter evolution; here, for Ising parameter ratio~$J/h=0.15$. Each error bar shows $\pm$standard error.
}
\label{fig:ising}
\end{figure*}

\begin{figure}
\centering
\includegraphics[width=0.45\textwidth]{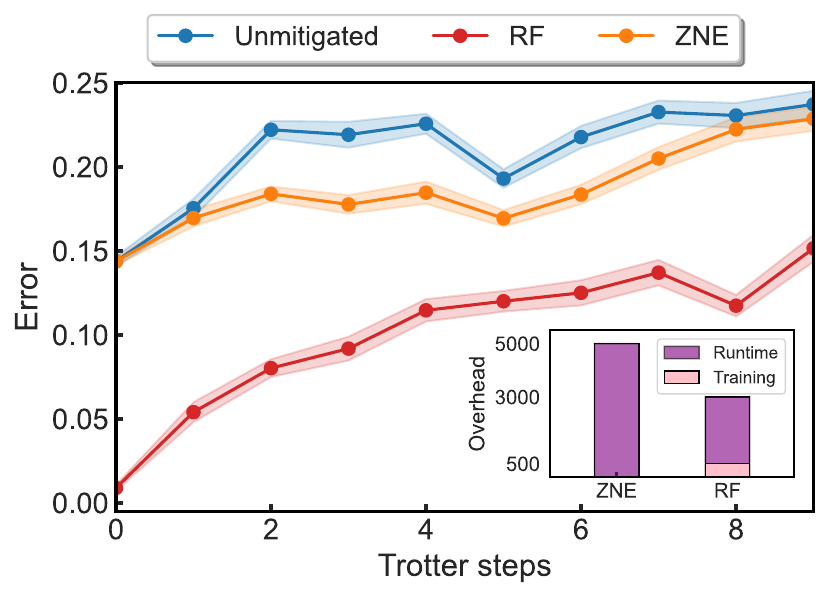}
\caption{
\textbf{Accuracy and overhead on QPU hardware for ML-QEM and QEM}
Average execution error of Trotter circuits for experiments on QPU device \texttt{ibm\_algiers} without mitigation and with ZNE  or ML-QEM (RF) mitigation. Error performance is averaged over $250$ Ising circuits per Trotter step, each with sampled Ising parameters~$J<h$ and each measured for all weight-one observables in a randomly chosen Pauli basis. Training is performed over $50$ circuits per Trotter step, which results in both a $40\%$ lower \textit{overall} and $50\%$ lower \textit{runtime} quantum resource overhead of RF compared to the overhead of the digital ZNE (see inset). Shaded regions represent $\pm$standard error.
}
\label{fig:ising_hardware}
\end{figure}

To benchmark the performance of the protocol on structured circuits, we consider Trotterized circuits. Here, we consider first-order Trotterized dynamics of the 1D transverse-field Ising model (TFIM) subject to different noise models based on the incoherent noise on the \texttt{FakeLima} simulator in Fig.~\ref{fig:ising}, before moving to experiments on IBM hardware with actual device noise in Fig.~\ref{fig:ising_hardware}. We observe that these circuits are not only broadly representative but also bear similarities to those used for Floquet dynamics \cite{Shtanko2023}. The dynamics of the spin chain is described by the Hamiltonian
\begin{equation*}
    \hat{H} = -J\sum_j \hat{Z}_j \hat{Z}_{j+1} + h\sum_j \hat{X}_j=-J\hat{H}_{ZZ}+h\hat{H}_{X}\;,
\end{equation*}
where $J$ denotes the exchange coupling strength between neighboring spins and $h$ represents the transverse magnetic field strength, whose first-order Trotterized circuit is shown in the inset of Fig.~\ref{fig:ising}. We generate multiple instances of the problem with varying numbers of Trotter steps and coupling strengths, such that the coupling strengths of each circuit are uniformly sampled from the paramagnetic phase ($J<h$) by choice. There are $300$ training and $300$ testing circuits per Trotter step, and the training circuits cover Trotter steps up to $14$. Each circuit is measured in a randomly chosen Pauli basis for all the weight-one observables. We then train the ML-QEM models on the ideal and noisy expectation values obtained from these circuits and compare their performance with digital ZNE. During the testing phase, we consider both interpolation and extrapolation. In interpolation, we test on circuits with sampled coupling strength $J$ not included in training but with Trotter steps included in training. In extrapolation (out-of-distribution testing), we test on circuits with sampled coupling strength $J$ not included in training as well as with Trotter steps exceeding the maximal steps present in the training circuits. 

On the noisy simulator in Fig.~\ref{fig:ising}, for this problem with incoherent gate noise in the absence (left) or presence (right) of readout error, ML-QEM (using RF) performs better than ZNE. We present a comparison across all ML-QEM models in App.~\ref{app:additional_results}, such that the RF model demonstrates the best performance among all ML-QEM models both in interpolation and extrapolation, closely followed by the MLP, OLS, and GNN. We envision that ML-QEM can be used to improve the accuracy of noisy quantum computations for circuits with gate depths exceeding those included in the training set. 

On the right of Fig.~\ref{fig:ising}, we consider the same problem in the second study but simulate the sampled circuits with additional coherent errors. The added coherent errors are CNOT gate over-rotations with an average over-rotational angle of $0.04\pi$. 
During the testing phase, the testing circuits cover $14$ more steps in extrapolation up to Trotter step $29$. Under the influence of added coherent noise, the performance of ML-QEM and digital ZNE deteriorate compared to the previous study. However, in the extrapolation scenario, none of the models demonstrate effective mitigation. In practical applications, a combination of, e.g., dynamical decoupling and randomized compiling, which can suppress all coherent errors, could be applied to the test circuits prior to utilizing ML-QEM models. This approach effectively converts the noise into incoherent noise, enabling the ML-QEM methods to perform optimally in extrapolation.

We benchmark the performance of the ML-QEM model against digital ZNE on real quantum hardware, IBM's \texttt{ibm\_algiers}. In this experiment, we do not apply any error suppression or additional error mitigation
; thus, the experiment involves incoherent, coherent, and readout noise, with the results shown in Fig.~\ref{fig:ising_hardware}. We train the ML-QEM with RF on $50$ circuits and test it on $250$ circuits at each Trotter step. We observe that $50$ training circuits per step, totaling $500$ training circuits, suffices to have the model trained well. With this low train-test split ratio, the ML-QEM reduces quantum resource overhead compared to ZNE \textit{both overall and at runtime}---the reduction is as large as $40\%$ overall and $50\%$ at runtime (see App.~\ref{app:additional_results} for details). We remark that the overall overhead reduction can be arbitrarily close to the runtime overhead reduction if given an arbitrarily large testing dataset size.  Additionally, we observe that the RF significantly outperforms ZNE for all Trotter steps, demonstrating the efficacy of this approach under a realistic scenario.

\begin{figure*}
\centering
\includegraphics[width=0.89\textwidth]{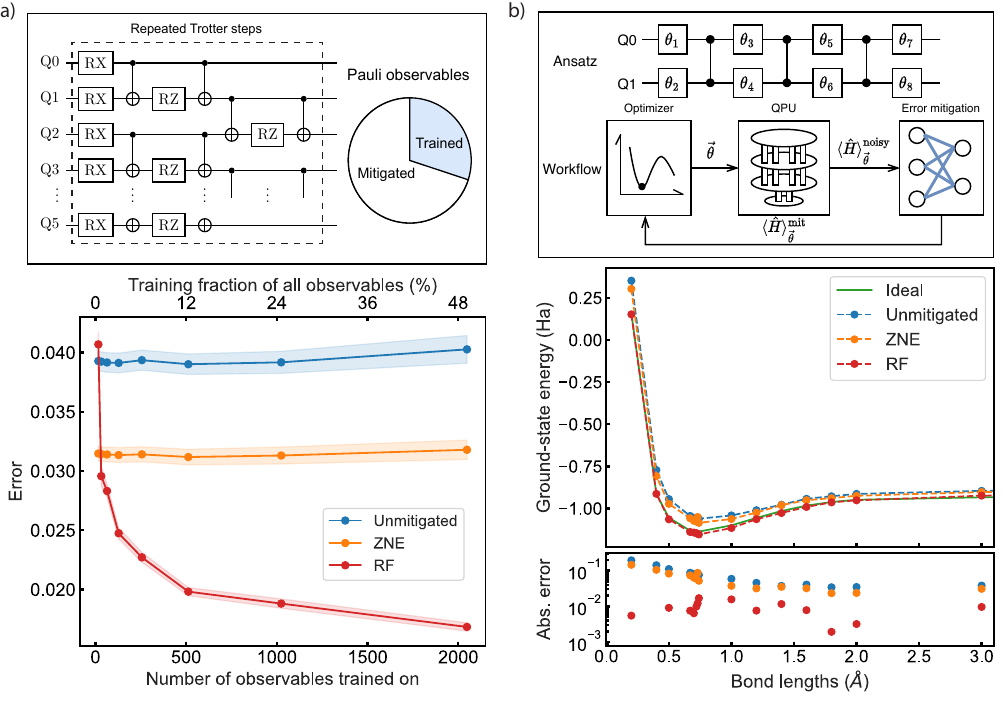}
\caption{
\textbf{Application of ML-QEM to a) unseen expectation values and b) the variational quantum eigensolver (VQE).}
a) 
Top: Schematic of a Trotter circuit, which prepares a many-body quantum state on $n=6$ qubits (in 5 Trotter steps). 
Top right: Circle depicts the pool of all possible  $4^n$ Pauli observables. Shadings depicts the fraction of observables used in training the ML model; the remaining observables are unseen prior to deployment in mitigation. 
Bottom: Average error of mitigated unseen Pauli observables versus the total number of distinct observables seen in training. Shaded regions represent the standard error.
 b) 
 Top: Schematic of the VQE ansatz circuit for 2 qubits parametrized by 8 angles $\vec{\theta}$. Below, a depiction of the VQE optimization workflow optimizing the set of angles $\vec\theta$ on a simulated QPU, yielding the noisy chemical energy $\langle \hat{H}\rangle_{\vec{\theta}}^\mathrm{noisy}$, which is first mitigated by the ML-QEM or QEM before being used in the optimizer as~$\langle \hat{H}\rangle_{\vec{\theta}}^\mathrm{mit}$. Compared to the ZNE method, the ML-QEM with RF method obviates the need for additional mitigation circuits at every optimization iteration at runtime. 
 Bottom: Ground energy estimate from VQE of the $\text{H}_2$ molecule as a function of the molecular bond lengths. The absolute error between the ZNE-mitigated and RF-mitigated estimates was calculated.
}
\label{fig:applications}
\end{figure*}

\begin{figure}[!t]
\centering
\includegraphics[width=0.46\textwidth]{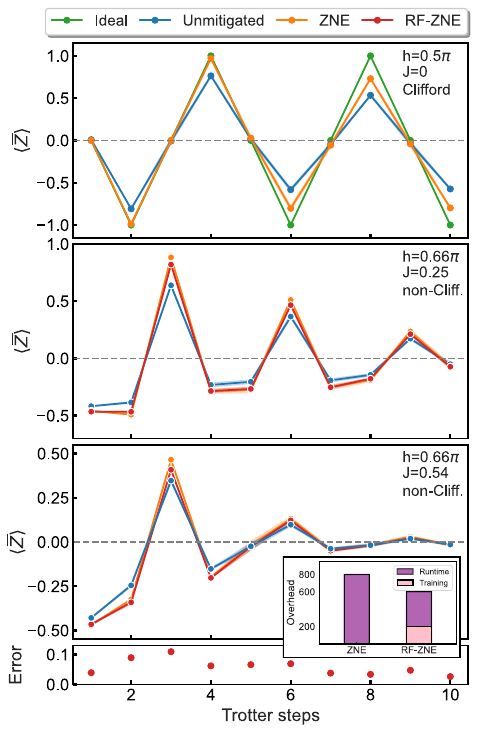}
\caption{
\textbf{ML-QEM mimicking QEM but with lower overheads on large, 100-qubit circuits executed on hardware.}
\textbf{Top three panels:} Average expectation values $\langle \hat{Z} \rangle$ from 100-qubit Trotterized 1D TFIM circuits run on the \texttt{ibm\_brisbane} QPU. Each panel corresponds to a distinct Ising parameter set, listed in the upper right. Top panel corresponds to a Clifford circuit, whose ideal, noise-free expectation values are represented by the green dots. The RF-mimicking-ZNE (RF-ZNE) curve corresponds to training the RF model on ZNE and readout-error-mitigated hardware data. This approach enables efficient, low-overhead error mitigation in scenarios where ideal outcomes from classical simulation are infeasible. Error bars (the shading) show $\pm$standard error.
\textbf{Bottom panel:} The $L_2$ norm of the error between ZNE-mitigated and RF-ZNE mitigated expectation values, averaged over non-Clifford testing circuits with randomly sampled couplings $J<h$. Each Trotter step represents an average over 40 test circuits, and the training includes 10 circuits per Trotter step. The approach achieves a $25\%$ reduction in \textit{overall} and $50\%$ reduction in runtime deployment in terms of quantum resource overhead compared to ZNE, as shown in the inset.
%
}
\label{fig:mimicry}
\end{figure}

\subsection{Mitigating Unseen Pauli Observables}\label{sec:applications_qst}

Routines, such as quantum state tomography and variational quantum eigensolver (VQE), require the measurement of multiple non-commuting Pauli observables on the same circuit, resulting in multiple target circuits with the same gate sequences but with different measuring basis. Error mitigation methods incur a large overhead by requiring additional mitigation circuits for each target circuit. Here, we show that it is possible to achieve better mitigation performance with lower overhead using ML-QEM.

In particular, we evaluate the performance of ML-QEM to mitigate unseen Pauli observables on a state $|\psi\rangle$ produced by the Trotterized Ising circuit depicted on the top of Fig.~\ref{fig:applications}(a), which contains $6$ qubits and 5 Trotter steps. We train the RF model on increasing fractions of the $4^6-1=4{,}095$ Pauli observables of a Trotterized Ising circuit with $J/h=0.15$, and then we apply the model to mitigate noisy expectation values from the \textit{rest} of all possible Pauli observables. The results of this study are plotted at the bottom of Fig.~\ref{fig:applications}(a). We observe that training the ML-QEM on just a small fraction $(\lesssim 2\%)$ of the Pauli observables results in mitigated expectation values with errors lower than when using ZNE.

\subsection{Enhancing Variational Algorithms}\label{sec:vqe}

The goal of the conventional VQE algorithm is to estimate the ground-state energy.. First, the energy $\langle \hat{H} \rangle_{\vec{\theta}}$ of the state prepared by a circuit ansatz $\hat{U}(\vec{\theta})$ with a fixed structure and parameters $\vec{\theta}$ is measured . Then, a classical optimizer is used to propose a new $\vec{\theta}$, and this procedure is executed repeatedly until $\langle \hat{H} \rangle_{\vec{\theta}}$ converges to its minimum. When executing this algorithm on a noisy quantum computer, error mitigation can be used to improve the noisy energy $\langle \hat{H} \rangle^\mathrm{noisy}_{\vec{\theta}}$ to the mitigated energy $\langle \hat{H}\rangle^\mathrm{mit}_{\vec{\theta}}$ and better estimate the ground-state energy. This workflow is shown at the top of Fig.~\ref{fig:applications}(b). Error-mitigated VQE with traditional methods can be costly, however, as additional mitigation circuits must be executed during each iteration. We use ML-QEM instead, where a model is trained beforehand to mitigate the energy estimate of an ansatz $\hat{U}(\vec{\theta})$, so that no additional mitigation circuits are executed each iteration, substantially lowering the runtime overhead. 

To demonstrate, we train the ML-QEM with RF on $2{,}000$ circuits with each parameter randomly sampled from $[-5, 5]$, and compute the dissociation curve of the H$_\mathrm{2}$ molecule on the bottom of Fig.~\ref{fig:applications}(b). The RF is trained on a two-local variational ansatz (depicted at the top of Fig.~\ref{fig:applications}(b)) across many randomly sampled $\{\vec{\theta}\}$. This method results in energies that are close to chemical accuracy. Notably, the absolute errors are smaller than those of the ZNE-mitigated energies. We remark that no readout error mitigation was applied here, and highlight the fact that ML-QEM, without any knowledge of the error model, can outperform a popular error mitigation method used naively.

\subsection{Scalability through Mimicry}\label{sec:mimicry}

For large-scale circuits whose ideal expectation values of certain observables are inefficient or impossible to obtain by classical simulations, we could train the model to mitigate expectation values towards values mitigated by \textit{other} QEM methods, enabling scalability of ML-QEM through mimicry. Mimicry can be visualized using the workflow for ML-QEM depicted in Fig.~\ref{fig:workflow} with an \textit{error-mitigated QPU} selected instead of a \textit{noiseless simulator}. Performing mimicry does not allow the ML-QEM model to outperform the mimicked QEM method, but reduces the overhead compared to the mimicked QEM.

We demonstrate this capability by training an ML-QEM model to mimic digital ZNE (together with twirled readout error extinction (TREX) mitigation~\cite{trex}) in a 100-qubit Trotterized 1D TFIM experiment on \texttt{ibm\_brisbane}. At utility scale, we emphasize that ML-QEM does not compete with any QEM techniques. Rather, its function is to accelerate conventional state-of-the-art methods by reducing their overhead through mimicry. Therefore, we choose ZNE as a widely-used QEM technique to be mimicked for this demonstration. 

We use ZNE to mitigate 5 single-qubit $\hat{Z}_i$ observables on 5 qubits on the Ising chain with varying numbers of Trotter steps and $J/h$ values. Each Trotter step contains $4$ layers of parallel CNOT gates, and the circuits at Trotter step $10$ has $1{,}980$ CNOT gates in total. As shown in the top panel of Fig.~\ref{fig:mimicry}, we first confirm that the ZNE-mitigated expectation values are more accurate than the unmitigated ones by benchmarking ZNE on a $100$-qubit Trotterized Ising circuit with $h=0.5\pi$ and $J=0$ such that it is Clifford and classically simulable. We then train a RF model to mitigate noisy expectation values in the same manner as ZNE. In this experiment, we apply Pauli twirling to all the circuits, each with $5$ twirls, before applying either extrapolation in digital ZNE or ML-QEM to mitigate the expectation values. 

We then find that the ML-QEM models accurately mimic the traditional method. The average distances, from the unmitigated result (after twirling average) to the mitigated expectation values produced by ZNE, and to the RF model mimicking ZNE, are close for all Trotter steps, as shown for specific $J$ and $h$ corresponding to non-Clifford circuits in the second and third panel of Fig.~\ref{fig:mimicry}. In the fourth (bottom) panel showing the residuals between the ZNE-mitigated and RF-mimicking-ZNE-mitigated values averaged over the training set comprising non-Clifford circuits, we see that RF mimics ZNE well. This result demonstrates that ML-QEM can scalably accelerate traditional quantum error mitigation methods by mimicking their behavior when ideal expectation values are classically intractable.

Importantly, this mimicry approach requires less quantum computational overhead \textit{both overall and at runtime}---here, $25\%$ lower \textit{overall} quantum computational resources and $50\%$ lower \textit{runtime} overhead (see App.~\ref{app:additional_results} for details).

\section{\label{sec:conclusion}Discussion}

In this paper, we have presented a comprehensive study of machine learning for quantum error mitigation (ML-QEM) methods for improving the accuracy of quantum computations. First, we conducted performance comparisons over many practically relevant contexts; they span unstructured and structured circuits, incoherent and coherent noise models, and applications (mitigating unseen Pauli observables and enhancing variational quantum eigensolvers). We find that the best-performing model is random forest regression (RF), pointing to the importance of selecting the appropriate model based on the complexity of the target quantum circuit and the desired level of error mitigation. Second, we demonstrated that ML-QEM can perform better than a traditional method, zero-noise extrapolation (ZNE). Paired with the ability to mitigate at runtime without additional mitigation circuits, ML-QEM reduces the runtime overhead. Therefore, ML-QEM can be especially useful for algorithms where many circuits that are similar to each other are executed repeatedly, such as quantum state tomography-like experiments and variational algorithms. Finally, we find that ML-QEM can even effectively mimic other mitigation methods, providing very similar performance but with a lower overall overhead. This allows ML-QEM to scale up to classically intractable circuits.

Future research in ML-QEM can focus on several directions to further enhance the training efficiency, performance, scalability, or generalizability of the method.  
First, better encoding strategies that incorporate other information about the circuit and the noise model, such as pulse shapes and output counts, may lead to even more accurate error mitigation. 
Second, the effect of the drift of noise in hardware can be studied to optimize the resources needed to adapt the ML-QEM models to drifted noise. As a step in this direction, we provide evidence in App.~\ref{app:trainig_efficiency} that the models can be efficiently fine-tuned to adapt to noise drifts. Third, ML-QEM can be benchmarked against other leading methods, such as PEC and probabilistic error amplification (PEA) \cite{Youngseok2023}. In this spirit, in App.~\ref{app:comparison}, we detail a resource analysis of these methods and the application of ML-QEM to them. Importantly, we underscore that ML-QEM does not itself compete with any of these methods. Rather, its function at utility scale is to accelerate conventional state-of-the-art methods by reducing their overhead through mimicry. Thus, the focus of ML-QEM and its future developments is to mimic the quality of state-of-the-art QEM approaches but with substantially less overhead in both training and execution runtime. 

In conclusion, our study underscores the potential of ML-QEM for enhancing quantum computations.

\section{Methods}

\subsection{Statistical Learning Models}\label{ssec:learning-based_methods}

Here, we discuss each of the statistical model (schematics shown in Extended Data Fig.~1), data encoding strategies, and hyperparameters used in this study. We emphasize that the performance of a model depends on factors such as the size of the training dataset, encoding scheme, model architecture, hyper-parameters, and particular tasks. Therefore, from a broader perspective, we hope that the models in this work provide a sufficient starting point for practitioners of quantum computation with noisy devices to make informed decisions about the most suitable approach for their application.

\begin{figure*}[!htbp]
\centering
\includegraphics[width=0.97\textwidth]{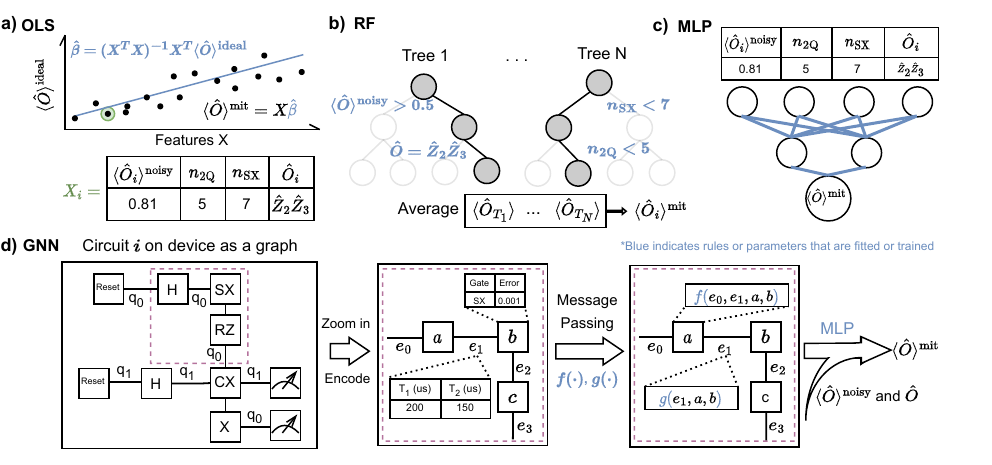}
\caption{
\textbf{Overview of the four ML-QEM models and their encoded features.} 
(a) Linear regression (specifically ordinary least-square (OLS)): input features are vectors including circuit features (such as the number of two-qubit gates $n_\mathrm{2Q}$ and SX gates $n_\mathrm{SX}$), noisy expectation values $\langle \hat{O} \rangle^\mathrm{noisy}$, and observables $\hat{O}$. The model consists of a linear function that maps input features to mitigated values $\langle \hat{O} \rangle^\mathrm{mit}$. (b) Random forest (RF): the model consists of an ensemble of decision trees and produces a prediction by averaging the predictions from each tree. (c) Multi-layer perception (MLP): the same encoding as that for linear regression is used, and the model consists of one or more fully connected layers of neurons. The non-linear activation functions enable the approximation of non-linear relationships. (d) Graph neural network (GNN): graph-structured input data is used, with node and edge features encoding quantum circuit and noise information. The model consists of multiple layers of message-passing operations, capturing both local and global information within the graph and enabling intricate relationships to be modeled.
}
\label{fig:models}
\end{figure*}

\subsubsection{Linear Regression}\label{sec:ols}

Linear regression is a simple and interpretable method for ML-QEM, where the relationship between dependent variables (the ideal expectation values) and independent variables (the features extracted from quantum circuits and the noisy expectation values) is modeled using a linear function.

One very nice relevant work in this area is Clifford data regression (CDR), proposed by Czarnik et al. \cite{Czarnik2021, Czarnik2022, Lowe_2021}. In CDR, the inputs are single target quantum circuit and a single particular observable of interest $\hat O$.  
In its canonical form, CDR is designed to substitute nearly all non-Clifford gates with Clifford gates. A small number of non-Clifford gates is retained to use as control knobs to modulate the output distribution of the expectation values for the $\langle \hat O  \rangle$ under the near-Clifford set of circuits. In this strategy, each of the near-Clifford circuits mirrors the structure of the target circuit. The replacement strategy is predicated on the idea that the computational demand of the classical simulations of the near-Clifford circuits escalates with the number of non-Clifford gates --- ideally kept well below fifty to manage classical simulation costs effectively. For a single observable $\hat O $, a batch of sufficiently many near-Clifford circuits is constructed and executed in simulation and in hardware in this way, resulting in a sets of expectation values $(X_\mathrm{ideal},X_\mathrm{noisy})$. This data is used to fit a linear regression between ideal and  noisy expectation values for a particular circuit and observable $\hat O$.  The target circuit is then executed, obtaining it's noisy expectation value, which can be corrected by the linear regression model. 

While the CDR method focuses on a single circuit and a single expectation value, the ML-QEM methods focuses on a large class of circuits, such as Floquet time dynamics under different depths, parameters, and under many different possible expectation values. Because of the challenge of constructing near-Clifford circuits to simultaneously satisfy constraints for many different observables, to our knowledge, CDR has not yet been generalized to  handle many expectation values for many different circuits at the same time. Thus, for our experiments on many-body quantum simulations, the cost of CDR would be prohibitively high, we target in ML-QEM the expectation values of interest, which typically include all weight-one and weight-two Pauli observables, and are thus a very large batch that scales polynomially in the system size. Thus, the spirit of CDR is at present to maximize the quality of a single circuit and a single observable, while that of ML-QEM is to minimize the cost of mitigating a very large class of circuits, observables, and parameters.

 In ML-QEM, the input circuit-level features are extracted from the quantum circuits: counts of each native gate on the backend (native parameterized gates are counted in binned angles), the Pauli observable in sparse Pauli operator representation, and optional device-specific noise parameters. We train a linear regression model that takes in these circuit-level features as well as the noisy expectation values, and predicts the ideal expectation values. The model minimizes the sum squared error between the mitigated and the ideal expectation values using a closed-form solution, which is named ordinary least squares (OLS). The linear regression model can also be trained using other methods, such as ridge regression, least absolute shrinkage and selection operator regression (LASSO), or elastic net. These methods differ in their regularization techniques, which can help prevent overfitting and improve model generalization if necessary. In our experiments, we use OLS for its simplicity and ease of interpretation. We note that standard feature selection procedures also help to further prevent overfitting and collinearity in practice.

\subsubsection{Random Forest}

Random forest regression is a robust, interpretable, non-linear decision tree-based model to perform QEM. As an ensemble learning method, it employs bootstrap aggregating to combine the results produced from many decision trees, which enhances prediction accuracy and mitigates overfitting. Moreover, each decision tree within the RF utilizes a random subset of features to minimize correlation between trees, further improving prediction accuracy. 

The input features to the random forest model are identical to those used for OLS (see Sec.~\ref{sec:ols}). We train a RF regressor with a specified large number of decision trees on the training data. Given all the features (circuit-level ones and the noisy expectation values), the RF model averages the predictions from all its decision trees to produce an estimate of the ideal expectation value. 

For RF, we used $100$ tree estimators for each observable. The tree construction process follows a top-down, recursive, and greedy approach, using the classification and regression trees (CART) algorithm. For the splitting criterion, we employ the mean squared error reduction for regressions. For each tree, at least $2$ samples are required to split an internal node, and $1$ feature is considered when looking for the best split.

\subsubsection{Multi-layer Perceptron}

Multi-layer perceptrons (MLPs), first explored in the context of QEM in Ref.~\cite{Kim2020}, are feedforward artificial neural networks composed of layers of nodes, with each layer fully connected to the subsequent one. Nodes within the hidden layers utilize non-linear activation functions, such as the rectified linear unit (ReLU), enabling the MLP to model non-linear relationships.

We construct MLPs with $2$ dense layers with a hidden size of $64$ and the ReLU activation function. The input features are identical to those employed in the random forest model. To train the MLP, we minimize the mean squared error between the predicted and true ideal expectation values, employing backpropagation to update the neurons. The batch size is $32$, and the optimizer used is Adam \cite{Kingma2015AdamAM} with an initial learning rate of $0.001$. In practice, regularization techniques like dropout or weight decay can be used to prevent overfitting if necessary. The MLP method demonstrates competitive performance in mitigating noisy expectation values, as evidenced by our experiments. However, it should be noted that MLPs are also susceptible to overfitting in this context.

\subsubsection{Graph Neural Network}

As the most complex model among the four, graph neural networks (GNNs) are designed to work with graph-structured data, such as social networks~\cite{Ying2018} and chemistry~\cite{Reiser2022}. They can capture both local and global information within a graph, making them highly expressive and capable of modeling intricate relationships. However, their increased complexity results in higher computational costs, and they may be more challenging to implement and interpret. 

A core aspect of our ML-QEM with GNN lies in data encoding, which consists of encoding quantum circuits, and device noise parameters into graph structures suitable for GNNs. Before data encoding, each quantum circuit is first transpiled into hardware-native gates that adhere to the quantum device's connectivity. To encode them for GNN, the transpiled circuit is converted into an undirected acyclic graph. In the graph, each edge signifies a qubit acted upon by some operation, while each node corresponds to an operation such as a gate. These nodes are assigned vectors containing information about the gate type, gate errors, as well as the coherence times and readout errors of the qubits on which the gate operates. Additional device and qubit characterizations, such as qubit crosstalk and idling period duration, can be encoded on the edge or node, although their inclusion was found not helpful for the problems studied.

The acyclic graph of a quantum circuit, serves as input to the transformer convolution layers of the GNN. These message-passing layers iteratively process and aggregate encoded vectors on neighboring nodes and connected edges to update the assigned vector on each node. This enables the exchange of information based on graph connectivity, facilitating the extraction of useful information from the nodes which are the gate sequence in our context. The output, along with the noisy expectation values, is passed through dense layers to generate a graph-level prediction, specifically the mitigated expectation values. As a result, after training the layers using backpropagation to minimize the mean squared error between the noisy and ideal expectation values, the GNN model learns to perform QEM.

For the GNN, we use $2$ multi-head Transformer convolution layers \cite{ijcai2021p214} and ASAPooling layers \cite{Ranjan2019ASAPAS} followed by $2$ dense layers with a hidden size of $128$. Dropouts are added to various layers. As with the MLP, the batch size is $32$, and the optimizer used is Adam \cite{Kingma2015AdamAM} with an initial learning rate of $0.001$.

We remark here a potential direction for future investigation using GNN models on large datasets of quantum circuits. When trained with noisy expectation values from different noise models, it is interesting to investigate if setting the noise parameters encoded in GNN, to the low-noise regime (e.g., setting encoded gate error close to zero and the encoded coherence times to some large value) allows the GNN to extrapolate to near noise-free expectation values without ever training on the ideal values, and thus providing potential advantages in both accuracy and efficiency \textit{at scale}.

\subsection{Zero-noise Extrapolation}\label{sec:zne_configuration}

We use zero-noise extrapolation with digital gate folding on $2$-qubit gates, noise factors of $\{1, 3\}$, and linear extrapolation implemented via Ref.~\cite{Rivero2022ZNE}.




\section*{Data availability}
We make the datasets used in this research accessible in \url{https://github.com/qiskit-community/ml-qem/tree/research}~\cite{our_repo}.

\section*{Code availability}
We make the source code for simulations and hardware experiments presented in this paper, as well as the model configurations, training and evaluation scripts, accessible in \url{https://github.com/qiskit-community/ml-qem/tree/research}~\cite{our_repo}.

\section*{Acknowledgements}
We thank Brian Quanz, Patrick Rall, Oles Shtanko, Roland de Putter, Kunal Sharma, Barbara Jones, Sona Najafi, Minh Tran, Sarah Sheldon, Thaddeus Pelligrini, Grace Harper, Vinay Tripathi, Antonio Mezzacapo, Christa Zoufal, Travis Scholten, Bryce Fuller, Swarnadeep Majumder, Youngseok Kim, Pedro Rivero, Will Bradbury, Nate Gruver, and Kristan Temme for valuable discussions. The bond distance-dependent Hamiltonian coefficients for computing the ground-state energy of the H$_2$ molecule using the hybrid quantum-classical VQE algorithm were shared by the authors of Ref.~\cite{Kandala2017}. The authors received no specific funding for this work.

\section*{Author Contributions}
I.S. initiated the project. H.L., I.S., and C.S. conducted experiments and wrote the application software. D.W., H.L., A.S., and Z.M. designed experiments. Z.M. guided and interpreted the study. The manuscript was written by H.L., D.W., and Z.M. All authors provided suggestions for the experiment, discussed the results, and contributed to the manuscript.

\section{Competing Interests}
The authors declare no competing interests.


\clearpage
\onecolumngrid

\section*{Supplementary Information}
\begin{appendix}

\section{Depolarizing Noise} \label{app:depolarizing_noise}
We show here that the ideal expectation values of an observable $\hat{O}$ linearly depend on its noisy expectation values when the noisy channel of the circuit consists of successive layers of depolarizing channels. This is more general than the result shown in \cite{Czarnik2021}.

Consider $l$ successive layers of unitaries each associated with a depolarizing channel with some rate $p_l$, the noisy circuit acting on the input $\rho$, $\mathcal{\tilde{C}}(\rho)$, is written as $\mathcal{\tilde{C}}(\rho)=\mathcal{E}_l(U_l\mathcal{E}_{l-1}(U_{l-1}\dots\mathcal{E}_1(U_1\rho U_1^\dagger) \dots U_{l-1}^\dagger) U_l^\dagger)$, where $\mathcal{E}_l(\rho)=(p_l/D)I+(1-p_l)\rho$.

It can be shown by induction that $\mathcal{\tilde{C}}(\rho)=(p(l)/D)I+(1-p(l))U_l\dots U_1\rho U_1^\dagger \dots U_l^\dagger$, where $p(l)=1-\Pi_{i=1}^l(1-p_i)$ as follows. Assuming for $l=k$, $\mathcal{\tilde{C}}(\rho)=(p(k)/D)I+(1-p(k))U_k\dots U_1\rho U_1^\dagger \dots U_k^\dagger$, then for $l=k+1$, we have $\mathcal{\tilde{C}}(\rho)= (p(k)/D)I+(1-p(k))[p_{k+1}I/D+(1-p_{k+1})U_k\dots U_1\rho U_1^\dagger \dots U_k^\dagger]= (p(k+1)/D)I+(1-p(k+1))U_k\dots U_1\rho U_1^\dagger \dots U_k^\dagger$. The induction completes with a trivial base case.

Therefore, the noisy expectation value of $\hat{O}$ becomes 
\begin{equation*}
\begin{split}
    \text{Tr}(\tilde{\mathcal{C}}(\rho)\hat{O}) &= \frac{p(l)}{D}\text{Tr}(\hat{O})+(1-p(l))\text{Tr}(U_l \dots U_1\rho U_1^\dagger \dots U_l^\dagger\hat{O})\\
    &=\frac{p(l)}{D}\text{Tr}(\hat{O})+(1-p(l))\text{Tr}(\mathcal{C}(\rho)\hat{O})\;,
\end{split}
\end{equation*}
where $\text{Tr}(\mathcal{C}(\rho)\hat{O})$ is the ideal expectation value of $\hat{O}$. 

For Trotterized circuits with a fixed Trotter step and a fixed brickwork structure, the number of layers $l$ of unitaries in the circuit is also fixed. Assuming some fixed-rate depolarizing channels associated with the $l$ layers of unitaries, the noisy and ideal expectation values of some $\hat{O}$ on these Trotterized circuits with different parameters then lie on a line. Therefore, the ML-QEM method can mitigate the expectation values by linear regression from the noisy expectation values to the ideal ones, and the linear regression parameters can be learned to vary according to the number of layers $l$. The ML-QEM is thus \textit{unbiased} in this case. We note that ZNE with linear extrapolation is still \textit{biased} in this case, since the noise amplification effectively results in a different combined depolarizing rate $p'(l)=1-\Pi_{i=1}^{l}(1-p'_i)$, which leads to expectation values with differently amplified noises each lying on a different line towards the ideal expectation value, and thus the linear extrapolation cannot yield unbiased estimates. 

\section{Break-even in Overall Overhead of ML-QEM}

Assuming the mimicked QEM requires $m$ total executions of either the mitigation circuits or the circuit of interest (e.g., digital/analog ZNE usually has $m=2$ or $3$ noise factors), the total cost of the mimicked QEM, namely its runtime cost, is $mn_{\text{test}}$. The total cost, including training, for the ML model is $mn_{\text{train}} + n_{\text{test}}$. Equating these two yields the break-even train-test split ratio in the total cost of our mimicry compared to the traditional QEM: $n_{\text{train}}/n_{\text{test}}=(m-1)/m$. Our mimicry shows a higher overall efficiency when the train-test split ratio is smaller than $(m-1)/m$.

\section{Additional Experimental Details}
\label{app:additional_results}

All non-ideal expectation values in simulations and experiments presented in this paper are obtained from corresponding measurement statistics with $10{,}000$ shots. 

\subsection{Errors Used in Simulations}
We included the error parameters used in all simulations in Tab.~\ref{tab:error_rates}. They represent typical to high-end error strengths in devices. The coherent CNOT error, injected as a CNOT over-rotational angle, is not native to the fake backends (representative of real devices using measured and calibrated model and noise parameters), and was added in cases where we tried to investigate the performance of the ML models under relatively strong coherent errors.

\begin{table}[!hbt]
\centering
\caption{\textbf{Average error parameters used in simulations.} The error properties of the two fake backends used in the paper are averaged over the five qubits in each fake backend. Gate errors are reported in terms of average gate infidelity. For CNOT gates, we report contributions to the total average gate infidelity from incoherent and coherent errors.}
\begin{tabular}{llcc}
\hline
& & \texttt{FakeLima} & \texttt{FakeBelem} \\
\hline
\hline
 \multirow{3}{*}{Qubit} & $T_1$ &  $\SI{61}{\micro\second}$ & $\SI{80}{\micro\second}$\\
 & $T_2$ &  $\SI{73}{\micro\second}$ & $\SI{79}{\micro\second}$\\
 & Readout & $3.4\times 10^{-2}$ & $3.0\times 10^{-2}$\\
 \hline
 \multirow{3}{*}{Incoherent} & SX &  $4.4\times 10^{-4}$ & $4.1\times 10^{-4}$\\
 & X &  $4.4\times 10^{-4}$ & $4.1\times 10^{-4}$\\
 & CNOT&   $1.2\times 10^{-2}$ & $1.4\times 10^{-2}$\\
 \hline
 Coherent & CNOT &  $0.7\times 10^{-2}$ & $0.6\times 10^{-2}$\\
\hline
\end{tabular}
\label{tab:error_rates}
\end{table}

\subsection{Random Circuits}
In the study of 4-qubit random circuits presented in Sec.~\ref{ssec:random_cricuits}, we use the Qiskit function \texttt{qiskit.circuit.random.random\_circuit()} to generate the random circuits, which implements random sampling and placement of 1-qubit and 2-qubit gates, with randomly sampled parameters for any selected parametrized gates. The 2-qubit gate depth is measured after transpilation. We remark that the random circuits sampled at large depths may approximate the Haar distribution and have expectation values concentrated around $0$ to some extent \cite{Harrow_2009, Liao_2021}. We further report a p-value of around $6.7\times 10^{-11}$ under a one-tailed Z-test between the RF and ZNE results, showing statistical significance of ML-QEM outperforming ZNE.

Although a considerable level of parameter optimization was performed in this study, we emphasize that rigor hyperparameter optimization may impact the relative performance of these methods, and we leave this analysis to future work.

\begin{figure*}[!ht]
\centering
\includegraphics[width=0.97\textwidth]{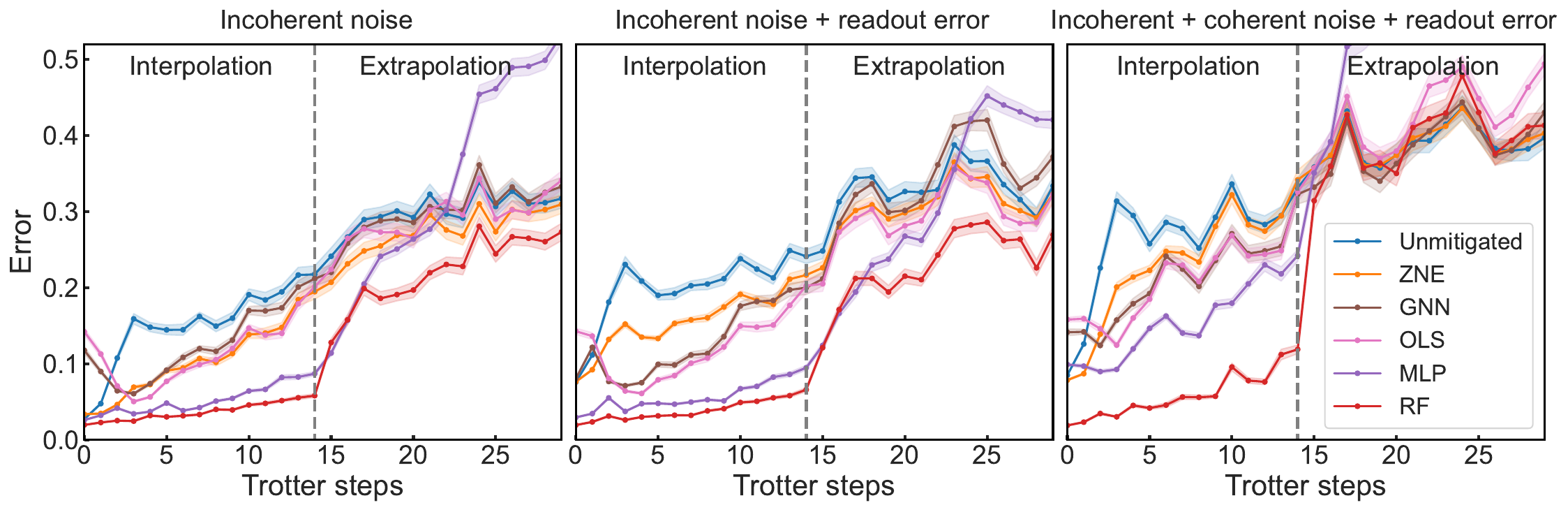}
\caption{
\textbf{ML-QEM and QEM performance for Trotter circuits.}
%
Expanded data corresponding to Fig.~\ref{fig:ising} of the main text that includes the three ML-QEM methods not shown earlier: GNN, OLS, MLP.
We study three noise models: Left: incoherent noise resembling \texttt{ibmq\_lima} without readout error, Middle: with the additional readout error, and Right: with the addition of coherent errors on the two-qubit CNOT gates. We show the depth-dependent performance of error mitigation averaged over $9{,}000$ Ising circuits, each with different coupling strengths $J$. 
For the incoherent noise model, all ML-QEM methods demonstrate improved performance even when mitigating circuits with depths larger than those included in the training set. However, all perform as poorly as the unmitigated case in extrapolation with additional coherent noise. Shaded regions represent the standard error.
}
\label{fig:ising_appendix}
\end{figure*}

\subsection{Trotterized 1D Transverse-field Ising Model}
In the study of the Trotterized 1D TFIM in Sec.~\ref{sec:comparison_ising}, we initialize the state devoid of spatial symmetries. This is done to intentionally introduce asymmetry in the single-qubit $\hat{Z}_i$ expectation value trajectories across Trotter steps, thereby increasing the difficulty of the regression task. Conversely, when the initial state possesses a certain degree of symmetry, the regression analysis, which incorporates noisy expectation values as features, becomes highly linear, resulting in a strong performance by the OLS method.

Regarding the expected mediocre performance across all techniques in the presence of coherent errors in extrapolation, as shown in Fig.~\ref{fig:ising_appendix}, we remark that coherent errors in
general introduce, to the leading order, quadratic changes to the expectation value as a function of depth,
versus incoherent errors introducing, to the leading order, linear changes~\cite{coherent_induces_quadratic}. These more rapid changes are certainly harder to model for ZNE which is designed to extrapolate around the vicinity of noisy expectation values, since they essentially make the noisy expectation values positioned arbitrarily between $–1$ and $+1$ at a given Trotter step. Hence no linear, quadratic, or exponential functions that are commonly used in ZNE extrapolation can consistently help bringing it close to the ideal expectation values. It is also reasonable that ML-QEM performs worse with stronger noises, and that significantly more training data may be needed for the model to capture the much larger variations in noisy expectation values induced by coherent errors in extrapolation.

We present a comparison across all ML-QEM models in the study of mitigating expectation values of Trotterized 1D TFIM in Fig.~\ref{fig:ising_appendix}. With incoherent noise and possible readout errors, the random forest model demonstrates the best performance among the ML-QEM models both in interpolation and extrapolation, closely followed by the MLP, OLS, and GNN. With additional coherent noise, in the interpolation scenario, the performance ranking of the other models remained largely consistent with that observed in the previous study. Notably, the random forest model exhibited the best performance among the ML-QEM models, closely followed by the MLP model. In the interpolation regime, we report the maximal p-values under the one-tail Z-test for RF outperforming ZNE: $1.8\times10^{-25}$ with incoherent noise, $6.3\times10^{-71}$ with added readout error, and $3.3\times10^{-44}$ with added coherent error; all indicate statistical significance. In the extrapolation regime, the maximal p-values for RF outperforming ZNE are $4.7\times10^{-2}$ with incoherent noise, and $5.9\times10^{-3}$ with added readout error.

For the experimental demonstration on real hardware \texttt{ibm\_algiers}, the ML-QEM requires $500 + 2{,}500=3{,}000$ total circuits, while running ZNE with $2$ noise factors on the testing circuits requires $2\times2{,}500 = 5{,}000$ total circuits. This means ML-QEM achieves a $40\%$ overall overhead reduction, and $50\%$ runtime overhead reduction. We report approximately $0.7$ QPU hours (based on a conservative sampling rate of $2~\si{\kilo\hertz}$ \cite{Youngseok2023}) to generate all the training data and seconds to train the RF model with a single-core laptop. We report a maximal p-value across all Trotter steps under the one-tail Z-test of $3.1\times10^{-10}$ for RF outperforming ZNE, indicating statistical significance.

As an aside, we observe that both in the simulation and in the experiment of the small-scale Trotterized 1D TFIM, there are significant correlations between the noisy expectation values and the ideal ones. There are also significant correlations but to a lesser degree between the gate counts and the ideal expectation values, suggesting the models are using certain depth information deduced from the gate counts to correct the noisy expectation values towards the ideal ones.

\subsection{Enhancing Variational Algorithms}
In the study presented in Sec.~\ref{sec:vqe}, when applying ML-QEM for enhancing variational algorithms such as VQE, the training overhead can be further lowered by taking advantage of the ability of ML-QEM to generalize to unseen Pauli observables (see Sec.~\ref{sec:applications_qst}) in the Hamiltonian. By decomposing $\hat{H}=\sum_i c_i \hat{P}_i$ into Pauli terms, the ML-QEM only needs to train on the sampled ansatz $\hat{U}(\vec{\theta})$ with a \textit{subset} of the Pauli observables in $\hat{H}$. This is demonstrated in our experiment shown in Fig.~\ref{fig:applications}(b) where the random forest is trained on sampled ansatz $\hat{U}(\theta)$ measured in $\hat{X}_1\hat{X}_2$ and $\hat{Z}_1\hat{Z}_2$ ($1{,}000$ for each observable), while the Hamiltonian of the $\text{H}_2$ molecule at each bond length consists of $\hat{X}_1\hat{X}_2$, $\hat{Z}_1\hat{Z}_2$, $\hat{I}_1\hat{Z}_2$ and $\hat{Z}_1\hat{I}_2$ Pauli observables~\cite{Kandala2017}.

\begin{figure}
\centering
\includegraphics[width=0.42\textwidth]{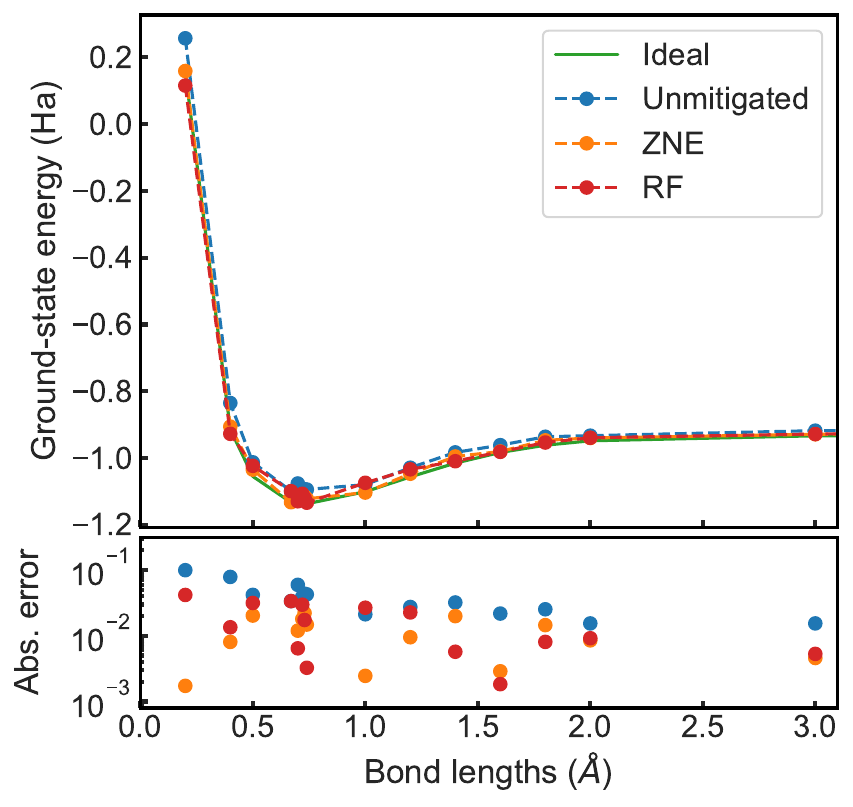}
\caption{
\textbf{VQE in the absence of readout errors.} We show the chemical energy $\langle\hat{H}\rangle_{\vec{\theta}}^\mathrm{noisy}$ of the $H_2$ molecule as a function of bond lengths, whose noisy estimate is first mitigated by the ML-QEM or QEM before being used in the optimizer as~$\langle \hat{H}\rangle_{\vec{\theta}}^\mathrm{mit}$. While ML-QEM can learn to mitigate the readout errors without the need for additional readout error mitigation, we explicitly show here the simulations in the absence of readout errors (essentially equivalent to applying state-of-the-art readout error mitigation techniques such as TREX). Absolute error is calculated between the ZNE-mitigated and RF-mitigated estimates. We observe closer-to-ideal noisy expectation values, and comparable performances between RF and ZNE mitigating the remaining errors. 
}
\label{fig:vqe_no_readout_err}
\end{figure}

To provide additional context, we also present the simulation of the same VQE task in the absence of readout errors, which is essentially the same as applying state-of-the-art readout error mitigation techniques such as TREX, as shown in Fig.~\ref{fig:vqe_no_readout_err}. All technical details are identical to those in the simulations in the presence of readout errors in Sec.~\ref{sec:vqe}. After removing readout errors, we observe closer-to-ideal noisy expectation values, indicating an easier task for QEM, and comparable performances between RF and ZNE mitigating the remaining errors. 

\subsection{Scalability through Mimicry}
In the study presented in Sec.~\ref{sec:mimicry}, we test on $40$ different coupling strengths $J$ for $h=0.66\pi$, each of which is used to generate $10$ circuits with up to $10$ Trotter steps, or $400$ test circuits in total. The traditional ZNE approach with 2 noise factors requires $2\times400=800$ circuits. In contrast, the RF-mimicking-ZNE approach here is trained with $10$ different coupling strengths $J$ for $h=0.66\pi$, each of which generates $10$ circuits with up to $10$ Trotter steps, or $100$ total training circuits. Therefore, the RF-mimicking-ZNE approach requires only $2\times100+400=600$ total circuits, resulting in $25\%$ \textit{overall} lower quantum computational resources. The savings are even more drastic \textit{at runtime}---again, the ZNE approach with $2$ noise factors requires $2$ circuits to be executed per test circuit, whereas each test circuit only has to be executed once for RF-mimicking-ZNE-based mitigation, resulting in $50\%$ savings.

We report approximately $0.14$ QPU hours(based on a conservative sampling rate of 2 kHz \cite{Youngseok2023}) to generate all the training data and seconds to train the RF model mimicking ZNE (RF-ZNE) with a single-core laptop for this experiment.

We remark that the testing (in-distribution) errors serve as an empirical error bound on the mitigation using the mimicry capability of ML-QEM. For instance, the error between RF-mimicking-ZNE and the ideal expectation values are upper bounded by approximately $0.1$, subject to the amount of training data we used in the mimicry experiment as shown in Fig.~\ref{fig:mimicry}. If the same RF model is to be used to mitigate similar circuits and observables, we can expect with high probability the error mitigation accuracy of the ML-QEM to be within $0.1$ error. 

Additionally, we remark that there are machine learning algorithms that can predict epistemic uncertainties by including learnable variance terms in the cost function~\cite{Kendall2017}. Such models can, in principle, output both the predicted value and the uncertainty in their prediction. It would be an interesting direction for future work to explore such learning algorithms in the context of QEM.

\section{Efficient Adaptability to Drifted Noise}
\label{app:trainig_efficiency}

\begin{figure}[!t]
\centering
\includegraphics[width=0.47\textwidth]{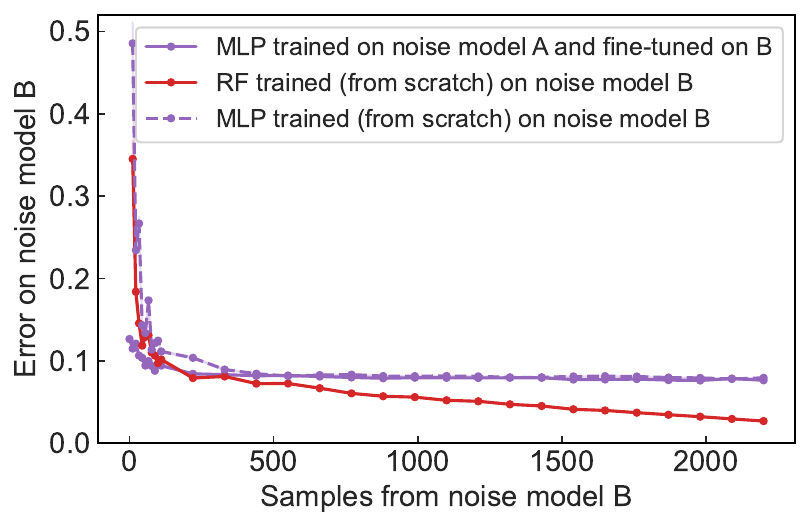}
\caption{
\textbf{Updating the ML-QEM models on the fly.}
Comparing the efficiency and performance of ML models, fine-tuned or trained from scratch, on a different noise model. Noise model A represents \texttt{FakeLima} and noise model B represents \texttt{FakeBelem}. All training, fine-tuning, and testing circuits are 4-qubit 1D TFIM measured in a random Pauli basis for four weight-one observables. The solid purple curve shows the testing error on noise model B of an MLP model originally trained on $2{,}200$ circuits run on noise model A and fine-tuned incrementally with circuits run on noise model B. The dashed purple curve shows the testing error on noise model B of another MLP model trained only on circuits from noise model B. The red curve shows the testing error on noise model B of an RF model trained only on circuits from noise model B. All three methods converge with a small number of training/fine-tuning samples from noise model B. While the testing error of the fine-tuned and trained-from-scratch MLP models converged, both were outperformed by a trained-from-scratch RF model. This provides evidence that ML-QEM can be efficient in training.
}
\label{fig:training_efficiency}
\end{figure}

Because the noise in quantum hardware can drift over time, an ML-QEM model trained on circuits run on a device at one point in time may not perform well at another point in time and may require adaption to the drifted noise model on the device. Therefore, we explore whether an ML-QEM model can be fine-tuned for a different noise model and show that similar performance can be achieved with substantially less training data.

In particular, we fine-tune an MLP and compare its learning rate against RF. The MLP can be fine-tuned on a different noise model after they have been originally trained on a noise model. The fine-tuning is expected to require only a small number of additional samples---this is demonstrated in Fig.~\ref{fig:training_efficiency} with the MLP trained on noise model A (\texttt{FakeLima}) and fine-tuned on noise model B (\texttt{FakeBelem}) which converges after $\sim300$ fine-tuning circuits. On the other hand, an MLP trained from scratch and tested on a noise model B shows a slower convergence after $\sim500$ training circuits, though both fine-tuning and training from scratch produce the same testing performance. We also compare them with an RF trained from scratch, which converges after fewer than $\sim300$ training circuits, demonstrating the excellent efficiency in training an RF model. While future research can investigate in more detail the drift in noise affecting the ML model performance, we show evidence that MLP can be efficiently adapted to new device noise and that RF can be trained as efficiently from scratch to new device noise.

\section{Comparison and Resource Analysis of QEM Methods}\label{app:comparison}

The quantum error mitigation (QEM) field features a rich variety of physics-based methods, including zero-noise extrapolation (ZNE), probabilistic error cancellation (PEC), probabilistic error amplification (PEA),  quantum subspace expansion (QSE), and Clifford data regression (CDR), each demonstrating notable efficacy under certain conditions and within particular scenarios~\cite{GoogleQEMReview2022,Temme2016, Li2017, McClean_2020, Czarnik2021, tsubouchi2023universal,Berg2022,Kim2023}. Our  machine learning-quantum error mitigation (ML-QEM) when used in the mimic setting could be trained on data from any of these methods. However some of these methods are more suitable to black-box and utility-scale problems than others. In this section, we briefly review salient features of these methods, their overall resource cost, and potential speedups offered by ML-QEM for these methods. Overall, we expect ML-QEM to perform well at mimicking all of these methods.

As discussed in the main text, we employ ML-QEM to mimic the effect of a physics-based quantum-error-mitigation (QEM) approach at utility-scale when classical numerics are intractable or unavailable. For the method of choice in our study,  we selected one of the most popular state-of-the-art QEM method in use today known as zero-noise extrapolation (ZNE). There are several variations of ZNE. We used the popular digital gate-folding variant of ZNE, which merits ease of use and easy cross-platform comparability.  Moreover, a key advantage of dZNE for our analysis is the simplicity in quantifying its sampling overhead, which is effectively the count number of noise-amplification factors utilized. This overhead is typically quite low for this QEM method.

In our study, we focus on developing an ML-QEM workflow designed for
broad applicability within a cloud-based quantum computing architecture.
This goal introduces several critical constraints on the workflow
to ensure its suitability and efficiency across a wide range of applications.
Firstly, the workflow must be sufficiently general to address quantum
circuits without requiring detailed knowledge of specific properties
or symmetries. Secondly, it should efficiently process numerous expectation
values, a necessity for general quantum computations. Lastly, both
the training and deployment phases must exhibit low runtime or sampling
overhead to remain feasible for widespread use.

In practical terms, our experiments trained ML-QEM to accommodate
a wide array of circuits, including those based on Trotter or Floquet
parameterization, spanning a diverse range of angle parameters and
two-qubit depths (up to $40$). Moreover, the experiments target a
large set of observables, often all weight one and weight two Pauli
operators. Essentially, ML-QEM strives for maximum breadth in its
application to various circuits, observables, and parameters.

Given these requirements, certain traditional QEM methods are more
suited to mimic and compare than others. Below, we explore the advantages
and current limitations of several prominent QEM strategies. We present
a comparative analysis of their precision and overhead in Tab.~\ref{tab:comparison},
juxtaposing them with the broad capabilities of ML-QEM and its ability
to enhance each of these method's runtime. Notably, we exclude quantum
subspace expansion (QSE) methods from our consideration, as they typically
cater to more specialized problems and rely on specific knowledge
about the problem, such as optimization objectives, or unique symmetries,
which does not
align with our criteria for general applicability and efficiency.
Additionally, we comment on the strategy of CDR, whose canonical
philosophy is the opposite of ML-QEM in that CDR aims for maximum
precision for one specific observable of one specific circuit. 

Before proceeding, we remark that in the expansive landscape of quantum error mitigation, it is now well-established that mitigating expectation values in an unbiased manner requires a superpolynomial number of samples in the worst-case scenario. This principle, developed across a series of papers, spans a wide array of error-mitigation techniques and is summarized in the review of Ref.~\cite{GoogleQEMReview2022}. Specifically, both ZNE and PEC necessitate an exponential number of samples relative to the number of gates within the effective light cone of the target observable. Crucially, the exponent of this relationship is intricately dependent on the noise level and the effective operator velocity. 
In the following, we provide a more quantitative analysis.

\begin{table*}[ht]
\centering
\caption{\textbf{Comparison of quantum error mitigation methods.} The ``*" rating for ML-QEM for mimickry means that the rating is the similar to that of the technique being mimicked. We assign one of three possible ratings to the accuracy, precision, scalability, runtime efficiency (training), and runtime efficient (mitigation) categories for each method based on their typical usage in literature: low, medium, and high. A high accuracy means that the method is, under realistic scenarios, unbiased; a high precision means that the error bar is of similar order to the shot noise; a high scalability means that the method has been demonstrated on quantum circuits with at least 100 qubits; a high (low) runtime efficiency in training means that the overhead is negligible (is a significant portion of the overall runtime overhead); and a high (medium) [low] runtime efficiency in mitigation means that the overhead is negligible (a small linear factor) [exponential].}
\begin{tabular}{lcccccc}
\hline
Method & General Obs. & Accuracy & Precision & Scalability & \multicolumn{2}{c}{Runtime efficiency} \\
\cline{6-7}
 & & & & & Training & Mitigation \\
\hline
\hline
Digital ZNE & Yes & Medium & Medium & High & High & Medium \\
PEC & Yes & High & High & Low & Low & Low \\
PEA & Yes & High & Medium & High & Low & Medium \\
QSE & No & High & High & Low & Medium & Low \\
\hline
ML-QEM (Mimickry) & Yes & * & * & High & Medium & High \\
\hline
\end{tabular}
\label{tab:comparison}
\end{table*}


\subsection{Probabilistic Error Cancellation (PEC)}

Probabilistic error cancellation (PEC) \cite{Temme2016,Berg2022}
is a state-of-the-art technique for mitigating quantum errors by constructing
an ensemble of of related quantum circuits to counteract the effects
of noise. It requires both a learning and an execution step. The learning
step depends on the input circuit---it can be fast for very structured
circuits or time consuming for circuits that have
many unique layers for reasons described in the following. Furthermore, the
runtime execution cost of PEC scales exponentially in the circuit
area $nl$, where $n$ is the number of qubits in the circuit and
$l$ is the depth of the circuit counted in two-qubit gate layers.
As we will explain, our experimental circuits used in this paper
appear to be well beyond the ability of PEC. 

The methodology of PEC is as follows: a quantum circuit $U$ is first
represented as a series of~$l$ sequential circuit layers $\left\{ U_{i}:i=1,\ldots l\right\} $,
each comprising parallelizable two-qubit native operations, such that
the target circuit is decomposed as $U=U_{l}\cdots U_{3}U_{2}U_{1}$.
Layer $U_{i}$ encapsulates the $i$-th consecutive set of operations.
The unique layers, totaling more than one but fewer than or equal
to $l$, form the basis for a preparatory learning phase, essential
for executing noise mitigation.

For each layer $i$, a comprehensive series of \emph{learning experiments}
is conducted to characterize the device's noise profile for the given
gate layer configuration. This characterization process hinges on
the application of Pauli twirling \cite{Bennett1996, Wallman2015, Minev2022}, which simplifies the noise into
a more manageable form, known as a stochastic Pauli channel. The same
twirling procedure is also used in the mitigation stage of PEC to
keep the noise in the learning and the mitigation as similar as possible.
In the learning, by employing an error amplification sequence---reminiscent
of those utilized in conventional device benchmarking---the required
data for constructing the noise model is gathered. These learning
circuits are iterated across various preparation and measurement bases,
as well as different amplification factors, to ensure thorough noise
characterization. However, the complexity and cost associated with
this procedure escalate with the device's connectivity and increase
linearly with the number of layers. The noise present in the device
is assumed to be of the form $\tilde{\mathcal{U}}_{i}:=\mathcal{U}_{i}\Lambda_{i}$,
with the ideal channel $\mathcal{U}_{i}\left(\rho\right)=U_{i}\rho U_{i}^{\dagger}$,
and the stochastic Pauli noise channel $\Lambda_{i}$. The data of
the learning experiments is used to fit a sparse model of $\Lambda_{i}$
that is most likely to produce the data. The following Pauli-Lindblad
form is assumed: 
\begin{equation}
\Lambda_{i}(\rho):=\exp[\mathcal{L}_{i}](\rho)\;,
\end{equation}
which is sparse in the Lindblad $\mathcal{L}_{i}$ generator terms,
defined according to the following canonical Lindblad form for Paulis
\begin{equation}
\mathcal{L}_{i}(\rho):=\sum_{k\in\mathcal{K}}\lambda_{k}(P_{k}\rho P_{k}^{\dagger}-\rho)\;.
\end{equation}
Here, $\mathcal{K}$ represents a select set of Pauli operators, $P_{k}$
from $\{I,X,Y,Z\}^{\otimes n}$, constrained by the device's assumed
noise-connectivity graph, and $\lambda_{k}$ are the sparse model
coefficients fit in the learning phase of PEC.

For the learning cost, it is worth noting that using a noise-generator
graph associated with $\mathcal{K}$ that has only a connectivity
degree of three, reflective of that of the physical layout of IBM
devices, has been observed to be practically effective. This sparse graph topology requires only nine bases
for the preparation and measurements of the learning protocol circuits,
substantially reducing the cost of the learning per layer. Still,
achieving satisfactory multiplicative precision for the noise parameters
$\lambda_{k}$ necessitates multiple points of noise amplification
and the PEC learning circuits are a non-negligible cost in general.
Moreover, it is worth noting that there is a significant constraint
in the learning given by a no-go theorem  \cite{Chen2022c} that
limits the precision attainable for certain Pauli coefficient pairs,
permitting only their product to be determined with multiplicative
precision. Breaking this degeneracy introduce some susceptibility
to SPAM and some additive errors into the learning. However, at the
scale of current experiments, the protocol nonetheless appears to
work well in practice, despite some of these non-idealities. 

We now focus on the mitigation overhead based on the rigorous bound  derived in Ref.~\cite{Berg2022}, formulated through  essentially
a Chernoff-Hoeffding inequality. This bound ensures, with a probability
of $1-\delta'$, that the deviation from the ideal expectation value
of an operator $A$ to be measured for the final quantum state of
the circuit $\langle A\rangle_{\mathrm{ideal}}:=\mathrm{Tr}[AU\ketbra{0}{0}U^{\dagger}]$,
is bounded as follows: 
\begin{equation}
\label{eq:pec-cost-full}
\left|\langle A\rangle_{\mathrm{ideal}}-\langle A_{N}\rangle\right|\le(C^{l\tau}-1)+\gamma(l)\sqrt{\frac{2\log(2/\delta')}{N}}\;,
\end{equation}
where $C^{l\tau}$ is a scalar that encapsulates the bias due to errors
in noise learning, $N$ represents the number of experimental shots,
and $\gamma(l)$, the cumulative sampling overhead, is the product
of individual layer overheads $\gamma_{i}$: 
\[
\gamma_{i}=\exp\left(\sum_{k\in\mathcal{K}}2\lambda_{k,i}\right)\;.
\]
To simplify, let us assume error-free noise model learning (i.e.,
$C^{l\tau}=1$), thereby eliminating the bias term. Under this assumption,
and defining a maximum allowable error~$\epsilon$ between ideal
and mitigated expectation values, the minimum number of shots $N$
required for a user to sample is 
\begin{equation}
N=2\log\left(\frac{2}{\delta}\right)\left(\frac{\gamma(l)}{\epsilon}\right)^{2}\approx4\left(\frac{\gamma(l)}{\epsilon}\right)^{2}\;,
\end{equation}
assuming $\delta\approx0.01$. This formulation underscores the critical
role of $\gamma(l)$ in determining the experimental workload.

To generalize across different devices, we introduce the average sampling
overhead per layer per qubit $\bar{\gamma}$, reflecting device quality.
Expressing total sampling overhead as $\gamma(l)^{2}=\bar{\gamma}^{nd}$,
where $n$ is the number of qubits (including idle ones) per layer,
allows us to estimate the runtime for PEC error mitigation more universally.
Given current superconducting device metrics, $\bar{\gamma}$ ranges
between $1.01$ and $1.04$.

\begin{figure}[ht]
\centering
\includegraphics[width=0.43\textwidth]{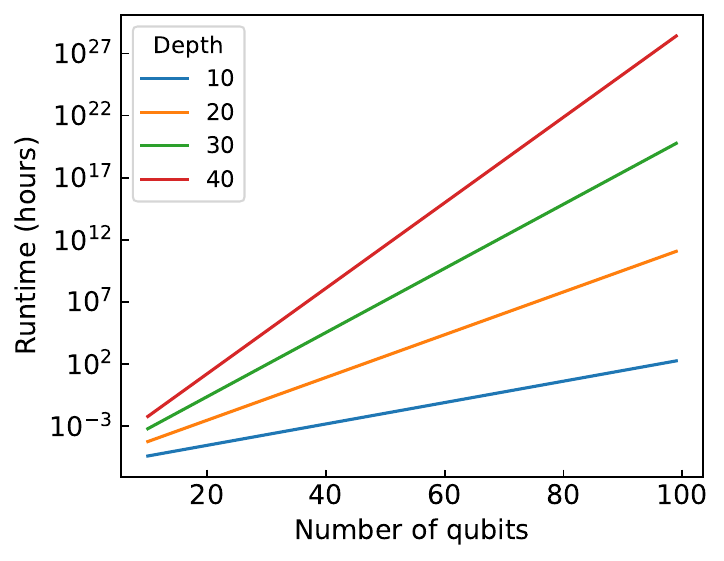}
\caption{\textbf{Runtime cost of probabilistic error cancellation (PEC) as a function of the width and depth of the mitigated quantum circuit}. 
The runtime cost is calculated from the sampling tail-bound, see Eqs.~\eqref{eq:pec-cost-full} and \eqref{eq:pec-cost-simple}, assuming a per-qubit sampling overhead~$\bar{\gamma}=1.02$ (a state-of-the-art number for today's machines) and a rate of  $\beta=1/5{,}000$ seconds per circuit layer operations (symbols defined in text). }
\label{fig:pec_runtime}
\end{figure}

Finally, combining this with the concept of circuit layer operations
per second (CLOPS) $\beta$, as introduced in Ref.~\cite{Wack2021},
provides a concise runtime cost expression for PEC error mitigation
across various devices and platforms,
\begin{equation}
\label{eq:pec-cost-simple}
\bar{\gamma}^{nl}\beta l\;,
\end{equation}
which is illustrated in Fig.~\ref{fig:pec_runtime}. This analysis reveals the exponential increase in runtime, making
PEC impractical for the large, deep circuits used in our experiments.
While techniques like lightcone reduction can significantly reduce
PEC sampling overhead, they do not yet render our deep circuits feasible
for PEC. Thus, for our experiments, we opted for zero-noise extrapolation
(ZNE), a more straightforward and currently popular method. 

\subsection{Zero-noise Extrapolation (ZNE)}

Zero-noise extrapolation (ZNE) \cite{Temme2016, Li2017}
is a popular error mitigation technique designed to infer noise-free
expectation values from data obtained from variations of a quantum
circuit under deliberately increased noise levels. The method extrapolates
these data to the (noise-free) zero-noise limit by assuming a relationship
between the noise amplification factor applied to the circuits and
the observed expectation values as a function of that noise. 

Unlike PEC, ZNE does not require explicit noise characterization or
calibration, operating under the primary assumption that the impact
of the noise on the expectation values scales predictably with the
amplified noise level. While the exact form of the dependence is not
known, it is typically assumed to have an analytical form. In practice,
one attempts a linear, polynomial, or exponential model---selected
based on heuristic or empirical considerations. Unlike PEC, ZNE is
not considered bias free, but due to its simplicity and experimental
effectiveness it is one of the most widely used methods \cite{Li2017}. 

While ZNE's operational simplicity facilitates its widespread use,
its theoretical framework does not in practice inherently provide
as tight bounds or precise estimations of sampling costs for achieving
a specified additive error $\epsilon$ with a confidence probability
$1-\delta$ as we would like, in contrast to what we observed more
closely with PEC. This complexity arises from both the indeterminate
functional form of the expectation value under unknown noise levels
and the challenge of calibrating the $M$ experimental noise amplification
factors $\{\lambda_{i}\}$. Generally, the cost of ZNE is anticipated
to scale exponentially with $M$, approximated loosely by $(2^{M}-1)^{2}$,
and due to the damping effect on expectation values as $\lambda$
increases, necessitating greater sampling accuracy for measuring increasingly
tiny signals. However, in practice, the method's cost is primarily
determined by the number of noise amplification factors, $M$, as
highlighted in the main text. Nonetheless, ZNE has demonstrated effective
performance in numerous experiments and is frequently applied in large-scale
quantum computations \cite{Kim2023,GoogleQEMReview2022,Youngseok2023}. 

Making the cavalier assumption that the expectation value signal decay
is exponential in the noise amplification factor, a simple analytical
cost expression can be derived \cite{Endo2019}. Consider
the exponential extrapolation based on two noise amplification factors
$\{\epsilon,r\epsilon\}$, where $\epsilon$ is the base noise level
of the target circuit and $r$ is the amplification scale for a single
noise amplification circuit. For an observable of interest $\hat{O}$,
the ZNE mitigated expectation value of $\hat{O}$, denoted as $\langle\hat{O}\rangle_{\mathrm{ZNE}}$,
can be inferred. If $\langle\hat{O}\rangle(\epsilon)$ and $\langle\hat{O}\rangle(r\epsilon)$
represent the measured expectation values at the corresponding noise
amplification levels of $1$ and $r$, respectively, the cost of this
extrapolation is approximated by: 
\[
\frac{r^{2}e^{2N\epsilon}+e^{2Nr\epsilon}}{(r-1)^{2}}\;,
\]
where $N$ is the total number of gates in the circuit. This approximation
assumes that $\langle\hat{O}\rangle(\epsilon)\propto\exp(-N\epsilon)$,
albeit this simplification may not capture the complexity of real-world
scenarios.

\subsection{Probabilistic Error Amplification (PEA)}

A main challenge for ZNE is the lack of a precise calibration of added
noise, leading to imprecision. This situation has recently been improved
experimentally by combining ZNE with the noise learning and error
injection of PEC into a method known as probabilistic error amplification
(PEA) \cite{Youngseok2023, Li2017}. The PEC--learned noise
model is well suited for amplifying the noise in ZNE in a well-controlled
and calibrated manner. Using the PEC notation introduced above, the
scaled the $i$-th noise channel $\Lambda_{i}^{1+\alpha}$, where
$\alpha$ is a tuning parameter, can be obtained from scaling the
Lindblad generator $\mathcal{L}$, introduced above, according to
$\exp\left(\alpha\mathcal{L}_{i}\right)$. This can be efficiently
implemented in practice using the standard quasi-probability (for
$\alpha<0$) or probability (for $\alpha>0$) sampling methods used
in PEC. The technique is very recent and new, but it appears to provide
a good middle ground between PEC and ZNE benefits and drawbacks. Based
on our experimental results with ZNE, we anticipate ML-QEM
to perform in a similarly efficient and effective manner with PEA as it
has with ZNE, though quantifying the overhead is more complicated due to the learning step.

\subsection{Clifford Data Regression (CDR)}

In Clifford Data Regression (CDR),
as proposed by Czarnik et al.~\cite{Czarnik2021,Czarnik2022,Lowe_2021},
the error mitigation task is focused on a singular quantum circuit
$U$ and a specific observable $\hat{O}$. At its core, CDR aims to
replace the majority of non-Clifford gates within $U$ with Clifford
gates, maintaining a minimal count of non-Clifford gates to prevent
the expectation values of $\hat{O}$ from uniformly converging to
zero, a common outcome for purely Clifford-based circuits.

The strategy behind this substitution rests on the understanding that
the computational overhead for classically simulating near-Clifford
circuits increases with the number of non-Clifford gates present.
To ensure practicality in classical simulation costs, the number of
non-Clifford gates is ideally kept under fifty. In practice, a set
of near-Clifford circuits is generated and both simulated and physically
executed to produce pairs of expectation values $(X_{\mathrm{ideal}},X_{\mathrm{noisy}})$.
This dataset facilitates the construction of a linear regression model
correlating ideal and noisy expectation values for the given circuit
$U$ and observable $\hat{O}$. Consequently, the noisy expectation
value obtained from executing the target circuit can be adjusted using
this model. When attempting to apply to multiple observables or circuits,
one inherent challenge for CDR is that naively each circuit and each
observable requires its own training data set, raising the cost of
execution. There is also an inherent challenge in crafting near-Clifford
circuits that meet the conflicting optimization requirements for multiple
observables concurrently to keep the training and deployment runtime
costs. In contrast, ML-QEM approaches aim to mitigate errors across
a broad spectrum of circuits and observables, including but not limited
to various configurations of Floquet time dynamics that encompass
a wide range of depths, parameters, and potential observables. In
effect, ML-QEM has the opposite spirit of CDR's canonical focus on
single-circuit, single-observable scenarios.

\subsection{Summary}

This section presented a comparative analysis of the learning and mitigation runtime overheads associated with several widely-used physics-based, black-box-style mitigation techniques, including zero-noise extrapolation (ZNE), probabilistic error cancellation (PEC), and probabilistic error amplification (PEA). Additionally, we explored closely related mitigation methods such as quantum subspace expansion (QSE) and Clifford data regression (CDR). Tab.~\ref{tab:comparison} provides a summary of the qualitative behaviors of the black-box methods.

Our experimental data showcase the utility-scale application of ML-QEM to ZNE mimicking. We explored the limitations and sampling overhead challenges presented by PEC. These analyses lend to our belief in the capabilities of ML-QEM, particularly its potential to significantly improve the runtime efficiency of these established and forthcoming physics-based mitigation strategies.

\end{appendix}

\bibliography{paper.bib}

\end{document}